\documentstyle[a4,12pt,epsf]{article}
\begin{document}
\author{V. Drchal$^{a,b}$, J. Kudrnovsk\'y$^{a,b}$, P. Bruno$^{c}$, 
P. Dederichs$^{d}$,\\
and P. Weinberger$^{b}$ \\
$^a$ Institute of Physics AS CR, CZ-180 40 Praha 8, Czech Republic\\
$^b$ Center for Computational Material Science, TU Vienna, Vienna, Austria\\
$^c$ MPI f\"ur Mikrostrukturphysik, D-06120 Halle, Germany\\
$^d$ Forschungszentrum J\"ulich, D-52425 J\"ulich, Germany\\
}
\title{The combined effect of temperature and disorder on interlayer 
exchange coupling in magnetic multilayers}
\date{}
\maketitle

\begin{abstract}

We study the combined effect of temperature and disorder in the
spacer on the interlayer exchange coupling.
The temperature dependence is treated on {\it ab initio} level.
We employ the spin-polarized surface Green function technique
within the tight-binding linear muffin-tin orbital method and
the Lloyd formulation of the IEC. 
The integrals involving the Fermi-Dirac distribution are calculated
using an efficient method based on representation of integrands by 
a sum of complex exponentials.
Application is made to Co/Cu$_{100-x}$M$_{x}$/Co(001) trilayers 
(M=Zn, Au, and Ni) with varying thicknesses of the spacer.
\end{abstract}

\section{INTRODUCTION}

The oscillatory interlayer exchange coupling (IEC) between magnetic
layers separated by a non-magnetic spacer has recently attracted 
considerable attention in the literature.
The physical origin of such oscillations is attributed to quantum
interferences due to spin-dependent confinement of electrons in the
spacer.
The periods of the oscillations with respect to the 
spacer thickness are determined by the spacer Fermi surface, and this
conclusion has been confirmed by numerous experiments.
In particular, the change of the Fermi surface by alloying
thus leads to the change of the oscillatory periods ({\it van Schilfgaarde
et al.}, 1995; {\it Kudrnovsk\'y et al.}, 1996).
On the other hand, there are very few studies of the temperature-dependence 
of the IEC ({\it Bruno}, 1995; {\it d'Albuquerque~e~Castro et al.}, 1996), 
and its systematic study on {\it ab initio} level is missing.

The main mechanism of the temperature dependence of the IEC is connected
with thermal excitations of electron-hole pairs across the Fermi level 
as described by the Fermi-Dirac function.
It turns out that other mechanisms (e.g. electron-phonon or electron-magnon
interactions) are less important.
The effect of the temperature on the IEC can be evaluated either
analytically or numerically.
The analytical approach assumes the limit of large spacer thicknesses, 
for which all the oscillatory contributions to the energy
integral cancel out with exception of those at the Fermi energy.
The energy integral is then evaluated by a standard saddle-point method 
({\it Bruno}, 1995).
The general functional form of the temperature-dependence of the interlayer 
exchange coupling ${\cal E}_x(T)$ in the limit of a single period is then:
\begin{equation} \label{eq_model}
{\cal E}_x(T) = {\cal E}_x(0) \, t(N,T) \, , \;\;\;\; 
t(N,T)=\frac{cNT}{\sinh(cNT)} \, .
\end{equation}
Here, $T$ denotes the temperature, $N$ is the spacer thickness in monolayers, 
and $c$ is the constant which depends on the spacer Fermi surface.
The term ${\cal E}_x(0)$ exhibits a standard $N^{-2}$-dependence 
({\it Bruno}, 1995),
while the scaling temperature factor $t(N,T)$ depends on $N$ via $NT$.
In the preasymptotic regime (small spacer thicknesses) the functional form
of $t(N,T)$ differs from that of Eq.~(\ref{eq_model}), particularly in the 
case of the complete but relatively weak confinement due to the rapid 
variation of the phase of the integrand which enters the evaluation of the 
IEC ({\it d'Albuquerque e Castro et al.}, 1996).

The second, numerical approach is in principle exact, not limited to large
spacer thicknesses, however, it may be numerically very demanding, in 
particular for low temperatures.
It is applicable also to disordered systems with randomness in
the spacer, magnetic layers, or at interfaces ({\it Bruno et al.}, 1996).

\section{FORMALISM}

The multilayer system consists of the left and right magnetic subspaces
separated by a non-magnetic spacer (the trilayer).
The spacer may be a random substitutional alloy.
We employ the  Lloyd formulation of the IEC combined with a
spin-polarized surface Green function technique as based
on the tight-binding linear muffin-tin orbital (TB-LMTO) method.
The exchange coupling energy ${\cal E}_x(T)$ can be written as 
\begin{eqnarray} \label{eq_IEC}
{\cal E}_x(T) &=& {\rm Im} \, I(T) \, , \quad
I(T) = \int_{C} f(T,z) \, \Psi(z) \, d z \, ,
\end{eqnarray}
where $f(T,z)$ is the Fermi-Dirac distribution function and
\begin{eqnarray} \label{eq_psi}
\Psi(z) &=& \frac{1}{\pi N_{\|}} \, \sum_{{\bf k}_{\|}} \,
{\rm tr}_{L} \, {\rm ln} \, {\sf M}({\bf k}_{\|},z) 
\end{eqnarray}
is a difference of (in the case of disorder, of configurationally 
averaged) grandcanonical potentials for the antiferromagnetic and 
ferromagnetic alignments of magnetic slabs ({\it Drchal et al.}, 1996).

The energy integration is performed over a contour $C$ along the real 
axis and closed by a large semicircle in the upper half of the complex 
energy plane, tr$_{L}$ denotes the trace over angular momentum indices 
$L=(\ell m)$, the sum runs over ${\bf k}_{\|}$-vectors in the surface 
Brillouin zone, and $N_{\|}$ is the number of lattice sites in one layer.
The quantity ${\sf M}({\bf k}_{\|},z)$ is expressed in terms of the
screened structure constants which couple neighboring (principal) layers
and of the so-called surface Green functions.
All details can be found in ({\it Drchal et al.}, 1996).
We only note that the use of a Green function formulation of the IEC 
is essential for describing the randomness in the spacer within the 
coherent potential approximation (CPA) which is known to reproduce 
compositional trends in random alloys reliably.

The integral in (\ref{eq_IEC}) can be recast into a more suitable form
using the analytic properties of $\Psi(z)$, namely, (i) $\Psi(z)$
is holomorphic in the upper complex halfplane, and (ii)
$z \Psi(z) \rightarrow 0$ for $z \rightarrow \infty, \, {\rm Im} z > 0$.
Let us define a new function $\Phi(y) = -i \, \Psi(E_F+iy)$ of a real
variable $y$, $y \geq 0$.
Then at $T=0$ K,
\begin{eqnarray} \label{eq_I0}
I(0) = \int_{0}^{+\infty} \Phi(y) \, dy \, ,
\end{eqnarray}
while at $T>0$ K,
\begin{eqnarray} \label{eq_IT}
I(T) = 2 \pi k_B T \sum_{k=1}^{\infty} \Phi(y_k) \, ,
\end{eqnarray}
where $k_B$ is the Boltzmann constant and $y_k$ are Matsubara energies
$y_k = \pi k_B T (2k - 1)$.
In the limit $T \rightarrow 0$, $I(T) \rightarrow I(0)$ continuously.

We have verified that the function $\Phi(y)$ can be represented with
a high accuracy as a sum of a few complex exponentials in the form
\begin{equation}
\Phi(y)=\sum_{j=1}^M \, A_j \, {\rm exp} (p_j y) \, ,
\label{dcexp}
\end{equation}
where $A_j$ are complex amplitudes and $p_j$ are complex wave numbers.
An efficient method of finding the parameters $A_j$ and $p_j$ is described
elsewhere ({\it Drchal et al.}, 1998).
The evaluation of $I(T)$ is then straightforward:
\begin{equation}
I(T) = - 2 \pi k_B T \, \sum_{j=1}^M \,
\frac{A_j}{{\rm exp} \, (\pi k_B T p_j) - {\rm exp} \, (-\pi k_B T p_j)} \, ,
\label{cexpT}
\end{equation}
which for $T=0$ K gives
\begin{equation}
I(0) = - \sum_{j=1}^M \, \frac{A_j}{p_j} \, .
\label{cexp0}
\end{equation}

\section{RESULTS AND DISCUSSION}

Numerical studies were performed for an ideal fcc(001) layer
stack of the spacer (Cu) and magnetic (Co) layers with the experimental
lattice spacing of fcc Cu.
The spacer layers can contain the impurities (Zn, Ni, and Au) which form
the substitutional alloy with spacer atoms.
Possible lattice and layer relaxations are neglected.
Alloying with Ni, Zn, or Au alters the electron concentration and, 
consequently, modifies the Fermi surface and thus, in turn, also the 
temperature dependence of the IEC.
The most obvious effect of the alloying, for $T=0$, is the change of
the periods of oscillations connected with the change of the corresponding
spanning vectors of the alloy Fermi surface ({\it Kudrnovsk\'y et al.},
1996).
The more subtle effect of the alloying is connected with damping of
electron states and relaxation of symmetry rules due to alloying.

To determine the parameters of complex exponentials (\ref{dcexp}), we have
evaluated the function $\Phi(y)$ at 40 Matsubara energies corresponding 
to $T=25$ K.
We have verified that the results depend weakly on the actual value of the
parameter $T$.
Special care was devoted to the Brillouin zone integration.
The efficiency of the present approach allows us to perform calculations
with a large number of ${\bf k}_{\|}$-points in the irreducible part of
the surface Brillouin zone (ISBZ).
Note also that such calculations have to be done only once and then the
evaluation of the IEC for any temperature is an easy task.
In particular, we employ typically 40000 ${\bf k}_{\|}$-points in the ISBZ 
for the first Matsubara energies close to the Fermi energy and
the number of ${\bf k}_{\|}$-points then progressively decreases for points
distant from the real axis.
The present calculations agree with the results of conventional calculations
({\it Drchal et al.}, 1996) but they are much more efficient numerically, 
in particular when calculations for many different temperatures are required.

\epsfsize=30cm
\vglue 10mm
\epsffile{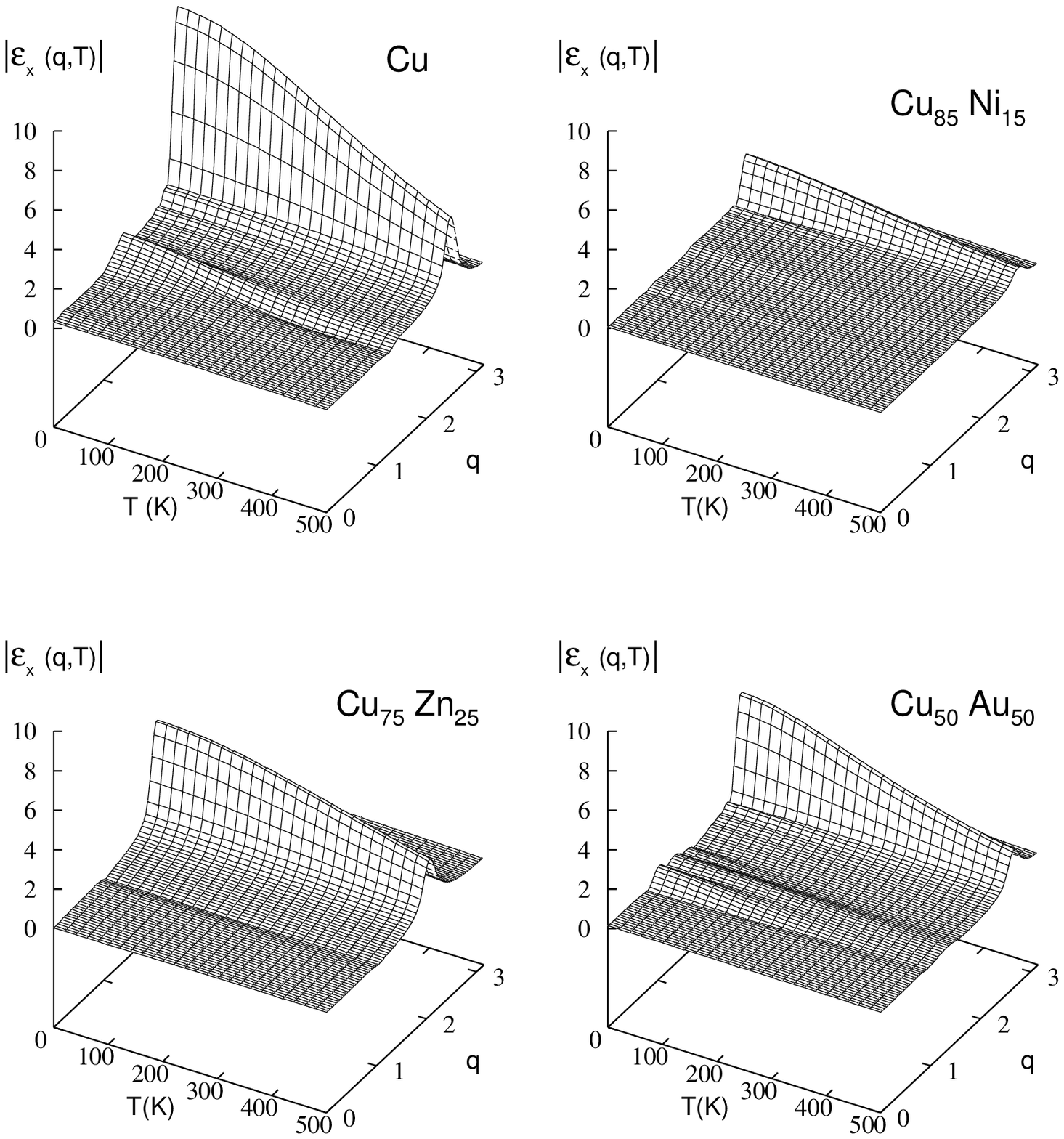}
\begin{figure} [htb]
\caption{ Absolute values of the discrete Fourier transformations of
   $N^2{\cal E}_x(N,T)$ with respect to the spacer thickness $N$ as
   a function of the temperature $T$ and the wave vector $q$ for a 
   trilayer consisting of semiinfinite Co-slabs sandwiching the spacer 
   with indicated compositions.}
\label{fig.1}
\end{figure}

The calculations were done for spacer thicknesses $N=1-50$ monolayers and 
for temperatures $T=0-500$~K (in steps 10~K) and by assuming semiinfinite 
Co-slabs.
In this case only one period, namely the so-called short-period exists, 
which simplifies the study.
There are several possibilities of how to present results 
(see {\it Drchal et al.}, 1998) for more details).
As an illustration, in Fig.~1 we plot the discrete Fourier transformations 
({\it Drchal et al.}, 1996) of $N^{2} \, {\cal E}_x(N,T)$ with respect 
to $N$, ${\cal E}_x(q,T)$, as a function of variables $q$ and $T$.
The discrete Fourier transformation on a subset $N \in 10-50$ which avoids 
the preasymptotic region is employed here.
The positions of peaks of $q=q_m$ then determine oscillation periods
$p=2 \pi/q_m$, while $|{\cal E}_x|$ give oscillation amplitudes 
({\it Drchal et al.}, 1996).
In particular, one can see how the modification of the Fermi surface
due to alloying changes the temperature dependence of the IEC, i.e.,
the coefficient $c$ in Eq.(\ref{eq_model}).

The following conclusions can be drawn from numerical results:
(i) The non-random case (Cu) exhibits the period $p \approx 2.53$~MLs
(monolayers) or, equivalently, $q_m \approx 2.48$ in accordance with
previous calculations ({\it Drchal et al.}, 1996). 
In accordance with ({\it Kudrnovsk\'y et al.}, 1996), the periods of 
oscillations for Cu$_{75}$Zn$_{25}$ alloy are shifted towards higher 
periods 
$(p \approx 3.05 MLs)$, towards smaller periods for Cu$_{85}$Ni$_{15}$
alloy $(p \approx 2.27 MLs)$, while they remain almost unchanged
for equiconcentration CuAu alloy spacer $(p \approx 2.36$~ MLs);
(ii) The periods of oscillations are temperature independent because
the electronic structure or, alternatively, spanning vectors are 
temperature independent;
(iii) The amplitudes exhibit a strong temperature dependence
in agreement with predictions of model theories ({\it Bruno}, 1995).
In particular, our results agree reasonably well with those of Fig.~3
in ({\it d'Albuquerque e Castro et al.}, 1996) for the case of ideal 
Cu spacer.
(iv) For alloy spacers at $T=0$ we mention, in particular, the dependence 
$N^{-2}$ of the oscillation amplitudes on the spacer thickness $N$ for 
CuNi and CuAu alloy spacers, while additional exponential damping due to 
disorder was found for the CuZn alloy.
This indicates a finite lifetime of states at the Fermi energy for
${\bf k}_{\|}$-vectors corresponding to short-period oscillations for this 
case and only a weak damping for CuAu and CuNi alloys;
(v) Finally, the effect of temperature (the factor $t(N,T)$ in 
Eq.~(\ref{eq_model})) is similar for a pure Cu-spacer, CuNi and CuAu alloys, 
but it is much smaller for the case of CuZn alloys.
The effect of temperature, similarly to alloying, is to broaden spanning 
vectors of the Fermi surface ({\it Bruno}, 1995).
If the damping due to alloying is non-negligible, the combined effect 
of disorder and temperature leads to a relatively smaller (compared to 
the case $T=$0~K) suppression of the oscillation amplitude with the 
temperature.

\bigskip \noindent {\bf Acknowledgment}

This work is a part of activities of the Center for Computational
Material Science sponsored by the Academy of Sciences of the Czech
Republic.
Financial support for this work was provided by the Grant Agency
of the Czech Republic (Project No. 202/97/0598), the Grant Agency
ot the Academy Sciences of the Czech Republic (Project A1010829), the 
Project 'Scientific and Technological Cooperation between Germany and 
the Czech Republic', the Center for the Computational Materials Science
in Vienna (GZ 45.422 and GZ 45.420), and the TMR Network 'Interface
Magnetism' of the European Commission (Contract No. EMRX-CT96-0089).

\bigskip \noindent {\bf References} \bigskip

\noindent {\it J.~d'Albuquerque e Castro, J.~Mathon, M.~Villeret,
                and A.~Umerski}, 1996, Phys. Rev. B {\bf 53}, R13306.

\noindent {\it P.~Bruno}, 1995, Phys. Rev. B {\bf 52}, 411.

\noindent {\it P.~Bruno, J.~Kudrnovsk\'y, V.~Drchal, and  I.~Turek},
               1996, Phys. Rev. Lett. {\bf 76}, 4254.

\noindent {\it V.~Drchal, J.~Kudrnovsk\'y, I.~Turek, and P.~Weinberger},
               1996, Phys. Rev. B {\bf 53}, 15036.

\noindent {\it V.~Drchal, J.~Kudrnovsk\'y, P.~Bruno, and P.~Weinberger},
               1998 (in preparation).

\noindent {\it J.~Kudrnovsk\'y, V.~Drchal, P.~Bruno, I.~Turek, and
               P.~Weinberger}, 1996, Phys. Rev. B {\bf 54}, R3738.

\noindent {\it M.~van~Schilfgaarde, F.~Herman, S.S.P.~Parkin, and
                J.~Kudrnovsk\'y}, 1995, Phys. Rev. Lett. {\bf 74}, 4063.

\end{document}